\documentstyle[12pt]{article}

\begin{document}

\title{A family of regular quantum interiors
for non-rotating black holes\, I: The GRNSS spacetimes}
\author{Emilio Elizalde$^{\cite{email,http}}$ and
Sergi R. Hildebrandt$^{ \cite{email2}}$
\\Instituto de Ciencias del Espacio (CSIC) \ \& \\
Institut d'Estudis Espacials de Catalunya (IEEC/CSIC) \\
Edifici Nexus, Gran Capit\`a 2-4, 08034 Barcelona, Spain}
\date{}
\maketitle


\newcommand{\rf}[1]{{\rm (\ref{#1})}}

\def\ab{\alpha \beta}
\def\lh{\hbox{{\boldmath \hbox{$ \ell $}}} }
\def\th{\hbox{{\boldmath \hbox{$ \Theta $}}} }      
\def\l{\Lambda}
\def\le{\Lambda_1} 
\def\li{\Lambda_2} 
\def\tr{{\tilde r} }
\def\dal{\sqcap_{\smash{\hskip -5.7pt \lower 1.3pt \hbox{--}}}\,\! }
\def\ks{Kerr-Schild }

\def\bce{\begin{center}}
\def\ece{\end{center}}
\def\beq{\begin{eqnarray}}
\def\eeq{\end{eqnarray}}
\def\ben{\begin{enumerate}}
\def\een{\end{enumerate}}
\def\ul{\underline}
\def\ni{\noindent}
\def\nn{\nonumber}
\def\bs{\bigskip}
\def\ms{\medskip}

\begin{abstract}
A seemingly natural mechanism is proposed, that could stop the
gravitational collapse of a very massive body. Without needing to
change the concept of the collapsing process itself, that is,
without invoking thin layers nor resorting to asymptoticity (as
has been usually done in the literature), it is proven that a
model can be built in which the quantum vacuum is able to produce
a negative stress that may stop the collapse of the black hole,
reaching a final state of the spacetime structure that is a static
de Sitter model. The solution is found by looking into a generic
family of spacetimes: that of maximal spherically symmetric ones
expanded by a geodesic radial null one-form from flat spacetime.
They are called here GRNSS spaces, and are proven to constitute a
distinguished family of \ks metrics. The models considered previously
in the literature are easily recovered in this approach, which
yields, moreover, an infinite set of possible candidates for the
interior of the black hole.
First steps towards their semiclassical
quantization are undertaken. It is shown that the quantization
protocol may be here more easily carried out than within conformal
field theory.
\end{abstract}

\newpage

\section{Introduction}
Soon after the appearance of General relativity, K. Schwarzschild
found a solution of Einstein's equations \cite{schw1} that was
useful to describe both the exterior and the interior spacetime
metric created by a spherically symmetric body in the case of
uniform energy density. This very early model already showed a new
feature of General Relativity that still concerns theoretical
physicists. According to General Relativity there exists a limit
involving the mass and the size of a star above which, in
accordance with classical matter properties, it cannot exist and
the appearance of a singularity is unavoidable
\cite{oppsny}--\cite{joshi}. In the late 60's, S.W. Hawking and R.
Penrose gave a general framework to analyze gravitational
collapse, and established their now well-known {\em Singularity
Theorems} (see e.g. \cite{he}). This seemed to settle down the
issue, but it has been suggested since then, for some decades,
that other effects, most likely those coming from quantum
mechanics ( see e.g. \cite{glinner}--\cite{markov1})
might alter those
predictions in a significant way (see e.g.
\cite{casimir}---\cite{eli1}). Nowadays, the issue of the
non-avoidance of singularities of space-time is still far from
being resolved. However, there are more and more indications that,
indeed, quantum effects may become most principal agents and that
a regularized object instead of a true singularity could be formed
after the collapse of a very massive object.

As everybody knows, Einstein's equations, or those corresponding
to the low energy limits of the usual candidates for suitable
quantum theories of gravitation, have two distinguished parts. One
corresponds to the geometry and the other to the stress-energy
content. It is clear that, for any solution to exist, the perfect
matching of both pieces is essential. Due to the difficulty in
getting (regularising) the contribution from the stress-energy
tensor, it is usually assumed that the spacetime can still be
described classically without much loss. In this sense, our work
(divided in two papers) gives a general solution for the geometry
that may be expected to be born after the collapse of a
non-rotating object takes place, following both old and recent
ideas on the type of core that is formed and on the precise way
this happens (see e.g. \cite{ip}--\cite{burinskii}). The case of
rotating objects will be considered elsewhere \cite{behm}.

The case dealt with in the present paper is only the first step of
the whole approach, corresponding to the resolution of the left
hand side of the aforementioned equations. Its purpose is to begin
the process of building up a way to find, within the distinguished
(and quite generic) family of solutions that will be obtained in
the first part of the paper, some compatible quantum fields that,
once regularized, may yield the same result as,
at least, a particular energy-momentum tensor inside this family.
As we shall see, the variety of possibilities that comes out, in
addition to those of some previous attempts by different authors
(also recovered here), and the good compliance with basic quantum
requirements, compels us to believe that this goal can be attained
and, in particular too, that black hole singularities might in the
end be indeed removed by quantum effects.

The present paper deals with the most plausible models, whereas in
a second one (referred to as paper II in what follows) it will be
proven that they constitute indeed the general solution for the
stress-energy that was assumed to exist. It is seen also, that
isotropization takes place far from the regularization scale. We
will also deal there with the possible extension of the formalism
to other candidates, e.g. stringy black holes.

This paper is two-fold. On one side, as mentioned elsewhere, we
show in detail how the obtaining of regularized sources for the
interiors of non-rotating black holes can be carried out, in fact,
without changing the initial concept of the collapsing process
itself, i.e. without having to invoke thin layers nor very
detailed asymptotic behaviors to get the expected results only
there. For the general case of an arbitrary equation of state, we
refer the reader to \cite{mps}, where some explicit solutions are
constructed, yet with a different type of energy-momentum tensor
than the one we consider here. In Sect.~\ref{s-fsus} we start with
a generic family of spacetimes, which is wide enough to include
all those that will be relevant for our purposes. In Sect.~\ref{s-gmc}, we 
write down the matching conditions
in order to couple any pair of spacetimes of the family under
study through an arbitrary hypersurface. We get the result that
the physical solution can only be a three-sphere of constant
``radius''. In Sect.~\ref{s-eqv}, we briefly review the
impossibility of matching Schwarzschild and de Sitter spacetimes
directly. In Sect.~\ref{s-csdse}, we specify the matching
conditions for a ``classical'' model of a Schwarzschild-de Sitter
exterior. And in Sect.~\ref{s-ri}, we write the correct solution
for the matching of a de Sitter interior, as the core of the
object, with an exterior which fits with current observational
data. One easily checks that the de Sitter core is the least
convergent regular solution, so that it is the limiting solution
between singularity-free and singular cores. In Sect.~\ref{s-i},
we prove that the transition expected to take place is
accomplished, provided we demand regularity of the collapsed body
only. We recover the two known models from our analysis, and show
that there is indeed an infinite set of possible interiors (see
also \cite{magli1}). In Sect.~\ref{s-sico}, we carry out a
numerical analysis for all the candidates of the preceding
section. The results show that the object is mostly far away from
the regularization scale where the notions of space and time lose
their sense, and therefore it still may be analyzed with some
approximate methods in order to find out possible quantum sources.
Before doing this, in Sect.~\ref{s-hirbh}, some aspects regarding
the horizons and the topological structure of the solutions are
briefly considered. In Sect.~\ref{s-ecs} we study the energy
conditions in these sources, and show that they can be easily
related to the properties of the energy density only. In
Sect.~\ref{s-qam}, we start the study of the sources that may
yield some of the solutions of the general family. In this
respect, we make the first steps towards their (semi-classical)
quantization. Preliminary results show that they are indeed good
candidates, deserving further analysis, being on the same footing
as when one deals with a conformal field theory.
At the same time, we show that they fulfill the conditions for
  an anomaly origin of their
sources. We also indicate how quantum field corrections can be
immediately incorporated into any initial classical model within
the family. A result which tells us, in particular, that
quantization may be carried out more easily here than in a
conformal quantum field theory. We end up with some final remarks. A brief 
survey can be found in \cite{rqibh}.
\section{The family of spacetimes under study}
\label{s-fsus} The spacetimes under study are constituted by the
family of maximal spherically symmetric spacetimes expanded by a
geodesic radial null one-form from flat spacetime (GRNSS spaces).
They are defined through the Kerr-Schild metric
\beq ds^2 = ds^2_{\eta} + 2 H(r) \lh \otimes \lh,
\eeq where $ ds^2_{\eta} $ stands for the
flat spacetime metric, $ H $ is an arbitrary function of $ r $
---the radial coordinate of the spherical symmetry--- defined in
some open region of the manifold, and $ \lh $ is a geodesic radial
null one-form. Another expression for the family is (recall that
the signature of the metric is ($-1 $, $ 1 $, $ 1 $, $ 1 $)) \beq
ds^2 = -(1-H) dt^2 + 2 H dt\,dr + (1+H) dr^2 + r^2 \bigl( d\theta^2 +
{\sin^2}\theta \, d\varphi^2 \bigr),
\eeq
where the coordinates are the customary spherical ones. In
these coordinates, $ \lh =  (1/\sqrt{2})(dt + dr) $. The other
possibility, i.e. $ \lh = (1/\sqrt{2})(dt - dr) $ yields the same
physical results. Clearly enough, all the metrics have spherical
symmetry and, since $ H=H(r) $, $ \partial_t $ is an
integrable Killing vector. In particular,  for $ H<1 $ it is
timelike, for $ H = 1 $ null, and for $ H > 1 $ spacelike. The
above coordinate choice avoids coordination problems near the
possible horizons and Cauchy hypersurfaces, e.g. when $ H = 1 $.

However it is worthwhile to make use of the existence of such
integrable Killing vector in order to write down the whole family
of metrics in a explicitly static form for the region where $ H <
1 $. The final expression is \beq
ds^2 = - (1-H){dt_s}^2 + {1 \over  1- H} dr^2 + r^2 \bigl( d\theta^2 +
{\sin^2}\theta \, d\varphi^2 \bigr),
\eeq
where $ dt_s $ is related to $ dt $ by
\beq
dt_s = dt - { H \over 1- H} dr .
\eeq This last expression for the family looks as a generalization
of the Schwarz\-schild metric and may help to identify the class
of spacetimes we are dealing with.\footnote{$ \lh $ can also be
written as $ \lh = (1/\sqrt{2})(dt_s+dr^*) $, where $ dr^* = { dr
/(1-H)}  $ is a direct generalization of the usual (Wheeler)
tortoise radial coordinate.}

The next task will be to characterize their geometrical
properties. To this end we will use an orthonormal cobasis (local
observer). In particular, we are going to choose the following,
which also avoids any problem at the possible horizons: \beq
\label{cobasis}
\th^0 = \biggl( 1- {H \over 2} \biggr) dt - {H \over 2}
dr, \quad \th^1 = \biggl( 1+{H \over 2} \biggr) dr + {H \over 2}
dt, \\ \th^2 = r \, d\theta, \quad \th^3 = r \sin \theta \,
d\varphi . \nn \eeq The Riemann tensor has the following non-zero
components, together with the ones obtained using index symmetry
(we follow the sign convention of \cite{mtw}) \beq
R_{0101} = - {{H''}\over 2} , \quad R_{0202} = R_{0303} = -{{H'} \over 2r}
, \quad R_{1212} = R_{1313} = {{H'} \over 2r} , \cr
R_{2323} = { H \over  r\sp{2}}.
\eeq

The Ricci tensor for these spacetimes has the following non-zero components
\beq
\label{ricci}
R_{00} = -R_{11} = -{1 \over 2}\left( H'' + {2 H'\over r} \right), \quad
R_{22} = R_{33} = {1 \over r}\Biggl( H' + {H \over r} \Biggr) .
\eeq
And the scalar curvature is given by
\beq
R = H'' + {4 H' \over r} + {2 H \over r\sp{2}} .
\eeq
Finally, the Einstein tensor has for components
\beq
\label{einstein}
G_{00} = -G_{11} = { 1 \over r} \Biggl( H' + { H \over r} \Biggr) ,
\quad G_{22} = G_{33} = - {1 \over 2}\left( H'' + {2 H'\over r} \right) .
\eeq
Whence, we see that the Einstein tensor satisfies the relations
\beq
G_{00} + G_{11} = 0, \quad G_{22} = G_{33}.
\eeq It is now clear that, imposing the additional relation $
G_{11} = G_{22} $, the whole set would be invariant under any
change of the cobasis, i.e. under any Lorentz transformation
(because in that case $ G_{\alpha \beta} $ would be proportional
to $ g_{\alpha \beta} $). In our situation, these relations are
invariant under {\it more restrictive} conditions. Actually under
any change of cobasis adapted to the spherical symmetry, i.e.
under any Lorentz transformation between $ \{ \th^0, \th^1 \} $
and between $ \{ \th^2, \th^3 \} $. As a consequence, the physical
interpretation of all these observers will be exactly the
same.\footnote{For instance, in the region with $ H < 1 $, where
the metric can be re-written in the explicit static form \beq
ds^2 = -(1-H){dt_s}^2 + (1- H)^{-1} dr^2 + r^2 ({d\theta}^2
\sin^2\theta
\, {d\varphi}^2 ),
\eeq
we have the natural, and customarily used, static
local observer associated with the cobasis $\th^0 = \sqrt{1-H}
dt_s, \th^1 = (1/ \sqrt{1-H}) d r, \th^2 = r d\theta, \th^3 = r
\sin \theta d\varphi $, to interpret the results.} This is the
reason why they are actually acceptable as consistent conditions
to represent an admissible generalization of the absolute
isotropic vacuum ($ G_{\alpha \beta} \propto g_{\alpha \beta}$)\footnote{The
spacetimes satisfying $ G_{00} + G_{11} = 0 $, and $ G_{22} = G_{33} $ are 
just
of the type being described here or a generalization of the Nariai
metric, see e.g. \cite{kramer}. However the latter cannot be
joined with any black hole model unless mass shells are
introduced, which is our aim to avoid, since they introduce
arbitrariness into the scheme. A detailed account will be given in
II.} ---see e.g. \cite{dymni,ds}.
Of course we could have imposed absolute isotropy in our
expressions. The result would then have been $ H (r) = A r^2 + B/r
$, where $ A $, $ B $ are arbitrary constants. One usually sets $
B = 0 $ in order to avoid singular behaviors near $ r = 0 $ (i.e. the 
Schwarzschild solution), thus
recovering the de Sitter spacetime for the isotropic vacuum, as
expected. However we have preferred to keep the metric in its more
general form. Actually  we will show below how a pure de Sitter
model is insufficient, for being considered as a reasonable model.
Notice, by the way, that the spherical vacuum conditions are here
obtained as a direct byproduct of the family of metrics under
study and they do not come from any specific imposition.

We shall use Einstein's field equations in the
form
\beq
\label{Einstein}
G_{\alpha \beta} = {8 \pi G_N \over c^2} T_{\alpha \beta} - \l
g_{\alpha \beta}.
\eeq
Usually, one chooses units for which $ G_N = 1 $, $ c= 1 $. If the
energy-matter content is of a quantum origin, the r.h.s. should be
understood as $ <{ T_{\ab}}> $, and $ <\l> $, in accordance with the
semiclassical approach to gravitation.
\section{General matching conditions}
\label{s-gmc} In this section we will write down the conditions
for two metrics of the preceding family to match with each other.
This kind of junction conditions constitutes by now a
well-established branch of geometry. However, there is no
canonical way of imposing them in the literature, and special care
has to be taken in order not to confuse the reader. We will follow
the formalism contained in \cite{israel2}--\cite{marsseno96}.

Our focus will be in matching two metrics with the aim of having
one of them
as an interior of the other, from some point in time onwards.
Here time refers to the time experienced by an observer
inside the static region of the spacetime manifold.

The general form of a hypersurface that clearly adjusts itself to the
spherical symmetry of any of these spacetimes is as follows
\beq
\Sigma : \cases{\theta = \lambda_{\theta}, \cr
\varphi = \lambda_{\varphi} , \cr r = r(\lambda) , \cr t =
t(\lambda) ,} \eeq where $ \{\lambda,\lambda_{\theta},
\lambda_{\varphi} \} $ are the parameters of the hypersurface. Its
tangent vectors are \beq
{\vec e}_{\theta} = \partial_{\lambda_{\theta}} \stackrel{\Sigma}{=}
\partial_{\theta}, \quad {\vec e}_{\varphi} = \partial_{\lambda_{\varphi}}
\stackrel{\Sigma}{=} \partial_{\varphi}, \quad {\vec e}_{\lambda} =
\partial_{\lambda} \stackrel{\Sigma}{=} {\dot r} \partial_{r}
+ {\dot t} \partial_t,
\eeq
where the dot means derivative with respect to $ \lambda $.
The normal one-form is then ($ {\bf n} \cdot {\vec e}_i = 0 $,
$ i = \{\theta, \varphi, \lambda\} $)
\beq
{\bf n} \stackrel{\Sigma}{=} \sigma({\dot r} dt - {\dot t} dr).
\eeq
$ \sigma $ is a free function in the case $ \bf n $ is a null one-form. 
Otherwise,
one can set $ {\bf n} \cdot {\bf n} = \pm 1 $, by chosing $ \sigma $ as
\beq
\sigma_{\pm} \stackrel{\Sigma}{=} {\pm 1 \over \sqrt{|{\dot t}^2-
{\dot r}^2+H ({\dot r} -{\dot t})^2}|}.
\eeq

The first junction conditions reduce to the coincidence of the
first differential form of $ \Sigma $ at each spacetime. We must
thus identify both hypersurfaces in some way.  Obviously, the
identification of $ (\lambda_i)_1 $ with $ (\lambda_i)_2 $ (1 and
2 label each of the spacetimes) is the most natural one, due to
the symmetry of the above scheme. This yields \beq \label{first1}
\Bigl[r(\lambda) \Bigr] = 0, \quad \hbox{where} \quad \bigl[ f
\bigr] \equiv (f_2 -f_1)_{\Sigma}, \\ \label{first2} \Bigl[{\dot
r}^2-{\dot t}^2+H ({\dot r} + {\dot t})^2\Bigr] = 0. \eeq

The second set of junction conditions comes from (we do not use
any mass shell, yet this could be added without problem) \beq
\Bigl[{\cal H}_{ij}\Bigr] = 0,
\eeq
where $ {\cal H}_{ij} $ is defined by
\beq
{\cal H}_{ij} \stackrel{\Sigma}{\equiv} - m_{\rho}\biggl({\partial^2 
\phi^{\rho}
\over \partial \lambda^i \partial \lambda^j} + { \hbox{\Large $
\Gamma $}}^{\rho}_{\mu \nu} {\partial \phi^{\mu} \over \partial
\lambda^i} {\partial \phi^{\nu} \over \partial \lambda^j} \biggr)
. \eeq
Here $ \vec{m} $ is a vector that completes the set $ \{
\vec e_{i} \} $ to form a vectorial basis of the
manifold.\footnote{For the cases in which $ \bf n $ is non-null, $
\vec m $ can be chosen simply as $ \vec n $, and $ {\cal H}_{ij} $ becomes 
the second fundamental form. $ {\cal H}_{ij} $ allows for dealing with a 
transition at the event horizon of the blak hole.} We will choose $$
\vec m = {\dot r} \partial_t - {\dot t} \partial_r . $$ This
choice has the property that $ \vec m \cdot {\bf n} = \sigma ({\dot
r}^2 + {\dot t}^2) $. Therefore, if it vanishes, the hypersurface,
$ \Sigma $, becomes degenerate and the joining process itself
cannot be carried out. Furthermore, $ \phi^{\rho} (\lambda) $ are
the parametric equations of the hypersurface ($ \{ \phi^0,
\allowbreak\phi^1,\phi^2, \phi^3 \} = \{t,r,\theta,
\varphi\}_{\Sigma}  $), and $ \Gamma^{\rho}_{\mu \nu} $ are the
connection coefficients.

Further calculations yield two supplementary conditions, namely
\beq
\label{second1}
\bigl[r {\dot t}\bigr] = 0,
\\
\label{second2}
\biggl[ H ( {\ddot t} {\dot t} - {\ddot r}{\dot r})
+ (1+H){\ddot r}{\dot t} + (1-H) {\ddot t}{\dot r}
-  H'\biggl({\dot r}^3 + {{\dot t}^3 \over 2} + {{\dot r}^2 {\dot t} \over 
2} \biggr) \biggr] =0 ,
\eeq
where $ H' $ is the derivative of $ H $ with respect to $ r $.

In order to properly close the junction, it is still necessary to
choose the signs of $ \sigma $ to be the same in both regions,
e.g. to impose a natural interior vs. exterior matching. In order
to be sure to deal with all possible candidates, one is forced to
look for all different solutions of hypersurfaces and spacetimes
in the preceding equations. In fact, it turns out from
Eqs.~\rf{first1}--\rf{second2} that the matching conditions
translate into\footnote{There is also the possibility of having $
t + r = {\rm const.} $, but this implies $ r = 0 $ for a finite $
t $, besides representing a geodesic null motion. Therefore they
cannot properly descibe the final static and regular interior
for the matter of a black hole.} $$ \bigl[ r \bigr] = 0, \quad
\bigl[ {\dot t} \bigr] = 0, \quad \bigl[ H \bigr] = 0, \quad
\bigl[ H' \bigr] = 0 . $$ On the other side, $ H $ is simply a
function of $ r $, so that the two last relations may be viewed as
a set of implicit relations defining $ r(\lambda) $ in terms of
the coefficients of $ H_1 $ and $ H_2 $. Thus, one arrives at the
following conclusion:

``{\em The only acceptable hypersurfaces fulfilling
the matching conditions, that preserve the spherical
symmetry, between two spacetimes of the GNRSS family,
are those satisfying $ r_1(\lambda) = r_2(\lambda) = R = const. $,
$ \dot t_1 = \dot t_2 $, $ [ H ] = [ H' ] = 0  $.}''

Without losing generality, one can choose $ t_1  = t_2 =
\lambda $, because of the global existence of the Killing vector $
\partial_t $. Moreover, we realize that the chosen coordinates are
privileged ones, in which the matching is explicitly $ C^1 $. The
hypersurface, $ \Sigma $, will be timelike, null or spacelike
according to $ H < 1$, $ H = 1 $ or $ H > 1 $, respectively.

Our specific aim in the rest of this work will be to investigate
if this situation can be actually met within our family of
solutions (GNRSS space) and, also, the physical reliability of the
necessary matching condition itself, including the value of $ H $
at the matching hypersurface.

Of course, other solutions for the exterior picture of the object,
e.g. asymptotically stopping, etc, may be viewed as refined
treatments of the whole process. The final results, however,
should coincide with the final (stopped) stage we will consider.
\section{The effect of the quantum vacuum. Israel's conditions.}
\label{s-eqv} Ever since the formulation of the idea that a vacuum
state should finish in a mode with an equation of state of the
type $ \rho + p  = 0 $, where $ \rho $ is the vacuum energy
density and $ p $ its pressure (or stress), it seemed natural to
try to match a Schwarzschild solution, acting as an exterior
metric, with a de Sitter solution, acting as the interior one. If
we had considered the junction conditions choosing $ \vec l = \vec
n $, as is often done, we would have obtained, instead of
Eqs.~\rf{second1} and \rf{second2}, the following ones, that
account for the continuity of the second fundamental form, $
K_{ij} $.\footnote{Notice that $ K_{ij} = {\cal H}_{ij} $ taking
$ \vec l = \vec n $.} \beq
\biggl[\sigma \Bigl\{ -{\dot t}+ H({\dot t} + {\dot r})\Bigr\} \biggr] = 0,
\\
\biggl[ \sigma\biggl\{ {\ddot r} {\dot t}- {\ddot t}{\dot r} + H'\biggl(
{\dot r}^3- {{\dot t}^3 \over 2} + {3 \over 2} {\dot r}^2\biggr) +
{H H' \over 2}[3({\dot r}^2{\dot t} - {\dot t}^2{\dot r})+ {\dot r}^3
+ {\dot t}^3] \biggr\} \biggr] =0 .
\eeq
Thus setting $ r_1 = r_2 = R $, $ R $ a constant,
and $ H_1 = 2 G_N m_1/ c^2 r $, $ H_2 = (\li/3)
r^2 $ (being $ m_1 $, $ \li $ constants with well-known
interpretations), we see that the only possibility in order to make
them compatible is
$ H_1 = H_2 =1 $. In principle this yields an
interesting relation between
the exterior mass and the quantum stress
\beq
\li = 6 G_N m_1/ c^2 R^3 .
\eeq However, if $ H_1 =H_2 =1 $, it is clear that the first
fundamental form becomes degenerate, i.e. $ ds^2_\Sigma =
R^2(d\lambda_{\theta}^2+ {\sin \lambda_{\theta}}^2
d\lambda_\varphi^2) $. Thus we do not have a hypersurface but
actually a {\it surface}. This is in fact a direct consequence of
the presence of a horizon (notice that we have $ H=1 $). Its null
character makes the above formalism insufficient, and one has to
apply the junction conditions on a null hypersurface, or better do
it once for all possibilities, as has been the case in the former
section. This point  went unnoticed in \cite{sz}, where the
authors claimed to have matched Schwarzschild and de Sitter
metrics at the horizon. In fact the true result eventually turns
out to be negative as well, i.e. Schwarzschild and de Sitter can
{\it never} be directly joined (this is clearly seen from the solution of 
the former section). This fact impelled other authors
to consider a massive layer (usually thin) in order to overcome
the problem \cite{ip,fmm}.

A more direct demonstration of this impossibility is
through Israel's conditions. Israel's conditions are
usually associated with the continuity of the Einstein tensor in a
matching situation. Because of Einstein's field equations, they
can also be interpreted as conditions on the stress-energy tensor
across the hypersurface. They are contained in the preceding ones
and have proven to be useful in getting part of the set of
restrictions in a very quick way.

In terms of the Einstein tensor, they read
\beq
\Bigl[n_{\rho} G^{\rho}_{\alpha} \Bigr] = 0.
\eeq In the exterior region, where we have the Schwarz\-schild
solution, the Einstein tensor is zero, therefore we must have \beq
n_{\rho} (G_2)^{\rho}_{\alpha} = 0,
\eeq
being $ { G}_2 $ de Sitter's Einstein tensor which satisfies
$ {G}_2 =- \li {g} $, so that we get
\beq \li {\bf n} =  0  \eeq
as a necessary condition. This clearly sets $ \li = 0 $ and hence a 
contradiction.
\section{The case of a Schwarzschild-de Sitter exterior}
\label{s-csdse} We recall that our aim was to find a regular
interior of the exterior Schwarz\-schild geometry that may encompass
some of the prevailing quantum ideas. However, it is not wrong to
consider as the exterior solution a Schwarz\-schild\---de Sitter
model. That is to say, an exterior that may also match
with recent observational evidence of a non-zero value for the
cosmological constant (see e.g. \cite{perl,ries}) as well as
uncharged black holes (e.g. \cite{astro}).\footnote{In fact, other more 
general exterior spacetimes
could be chosen within our family of spacetimes. For instance,
other type of quantum vacuum contributions, see  also
Sect.~\ref{sss-dm}, or other classical solutions, as the
Reissner-Nordstr{\"o}m one (see Sect.~\ref{s-ecs}). See also the Final
Remarks section.}. The exterior metric is then generated by ($ G_N
= c = 1$) \beq H_1 = {2m_1 \over r} + {\le \over 3} r^2 , \eeq
where $ m_1 $ is the usual Schwarzschild mass, and $ \le $
accounts for the cosmological constant. This solution is also
known as the Kottler-Trefftz solution, see
\cite{kottler,trefftz,kramer,tolman}. As $ r \to \infty $ the
spacetime tends to a de Sitter model of cosmological constant $
\le $ (see also \cite{dymni,ds}).

The junction conditions translate then into
\beq
H_2(R) = H_1( R) = {2m_1\over R} + {\le \over 3} R^2, \\
H'_2(R) = H'_1(R) = -{2 m_1 \over R^2} + 2 {\le \over 3} R . \eeq
\section{Regular interiors}
\label{s-ri} The preceding conditions are the ones to be imposed
in order to have a proper matching of both spacetimes. Moreover,
we will also focus on those interior solutions which are
everywhere regular. From the expressions of the Riemann tensor and
the metric, we see that this may only be accomplished if \beq
\label{eq-reg} H_2(0) = 0, \quad H'_2(0) = 0 .
\eeq Thus, we finally encounter four conditions in order to have a
regular interior solution. Two of them are imposed on the matching
hypersurface, while the other two are imposed at the origin of the
spherical symmetry. We shall thus expand $ H_2 $ at $ r - R $ or $
r $, depending on the situation, or at their corresponding
adimensional values $ \tr $ or $ \tr -1 $, where $ \tr $ = $ r/ R
$.

From now on, we will consider $ H_2 $ to be an analytic function
of the variable $ \tr $, a most natural hypothesis in view of the
regular character prescribed for the interior
solution.\footnote{It may be possible to demonstrate that
analyticity or some degree of differentiability is unavoidable,
since $ H_2 $ is just a function of one variable that is only
defined in a closed region, namely $ r\in [0,R] $, and along
the interval it is fixed to be finite. Otherwise, this
instance serves to perform a study of the possible interiors.} In
this case, the origin conditions tell us that \beq H_2 (\tr ) =
\sum_{n=2}^{\infty} {b}_n {\tr}^n . \eeq Now, one has to impose
the two other conditions. Obviously it is the same to consider $
H_2 (\tr ) $ or $ H_2 (\tr-1) $ in the whole procedure. However,
we will first work with $ H_2 (\tr- 1) $ in order to implement the
junction conditions directly. From the preceding
result, one immediately has \beq H_2 = \sum_{n=0}^{\infty} {a}_n
(\tr-1)^n, \eeq and the junction conditions tell us that \beq
{a}_0 = H_1(1) = {2m_1\over R} + {\le \over 3} R^2, \\
{a}_1 = H'_1(1) = -2\biggl({ m_1 \over R} +  {\le \over 3} R^2 \biggr),
\eeq
where $ H_1 (\tr ) = (2m_1/R)(1/\tr) + (\le R^2 / 3) \tr^2 $ and the
prime denotes now derivation with respect to $ \tr $.

The following step is to impose regularity of the solution 
---Eqs.~\rf{eq-reg}. We get
\beq \sum_{n=0}^{\infty}(-1)^n {a}_n = 0 \eeq
and
\beq \sum_{n=0}^{\infty}(-1)^n n {a}_n = 0, \eeq
which, by virtue of the matching conditions, yield
\beq \sum_{n=0}^{\infty}(-1)^n {a}_{n+2} = {\le \over 3}R^2 -{4m_1 \over R} 
\eeq
and
\beq \sum_{n=0}^{\infty}(-1)^n (n+2) {a}_{n+2} = 2 \biggl({\le \over 3}R^2
-{m_1 \over R} \biggr) . \eeq
It is clear that there are infinitely many possible candidates for
these interiors.
In the following sections we will analyze in  more detail the
properties that the big family of candidates share in common.

A final remark is in order. The conditions at the origin to make
the final solution ``everywhere regular'', aside from the matching
being also fulfilled, are important. For instance, $ H = A + B r
$, $ A $ and $ B $ constants, can immediately satisfy the junction
conditions, therefore the metric is $ C^1 $ in these coordinates.
However $ G_{11} $ and $ G_{22} $ yield $$\displaylines{ G_{11} =
-\biggl({A\over r^2} + {2B \over r} \biggr), \quad G_{22} = - { B
\over r} , } $$ which is {\it not} regular at the origin, for
whatever values of $ A $ or $ B $ different from zero. Of course,
we could afford to study within these models a bigger family which
also included singular models at the origin. In this work, as said 
elsewhere, we will
just focus on regular models, because of the interest in studying
their plausibility. The singular situation will be
considered elsewhere.
\section{Isotropization}
\label{s-i} To summarize, a basic property of any candidate to an
interior solution is its behavior at the origin. First, due to the
regularity conditions, all Riemann invariants turn out to be
finite at the origin, as well as the components of the Einstein
tensor, in accordance with previous works
\cite{markov1,fmm,ip,mps}. Moreover, it is also very important to
further analyze how they behave near the origin.

Taking into account the expression of $ H_2 $ in powers of $ \tr
$, we get \beq G_{11} =-{1\over R^2} \sum_{l=2}^{\infty} (l +1)
{b}_l {\tr}^{l-2}, \eeq and \beq G_{22} =-{1\over R^2}
\sum_{l=2}^{\infty} {l \choose 2}{b}_l {\tr}^{l-2}. \eeq It is
then clear that $ G_{11} $ and $ G_{22} $ are different from each
other.\footnote{\label{footi}It is important to notice that the
conservation of the Einstein tensor implies a relation between $
G_{11} $ and $ G_{22} $ of the type: $ G_{22} = G_{11} + G_{11}'
r/2 $. The same holds for the stress-energy tensor.}
Yet we have
the very relevant property that, for any of these spacetimes, it
holds \beq \lim_{\tr \to 0} G_{11} = \lim_{\tr \to 0}( -G_{00}) =
\lim_{\tr \to 0} G_{22} = \lim_{\tr \to 0} G_{33} = -{3 {b}_2
\over R^2} . \eeq Whence, we see that a general isotropization
{\it independent of the model} is actually accomplished in a
completely natural way. In terms of $ a_l $ we get \beq
G_{11} =-{1\over R^2} \sum_{M=0}^{\infty} {A}_M {\tr}^M, \\
{A}_M = (-1)^M (M+3) \sum_{l=M+2}^{\infty} (-1)^l {l \choose l-2-M}a_l , \nn
\eeq
and
\beq G_{22} =-{1\over R^2} \sum_{M = 0}^{\infty}{M+2 \over 2} {A}_M {\tr}^M. 
\eeq
So that
$$\displaylines{
\lim_{\tr \to 0} G_{11} = \lim_{\tr \to 0}( -G_{00}) = \lim_{\tr \to 0}
G_{22} = \lim_{\tr \to 0} G_{33} = -{ {A}_0 \over R^2}, \cr
{A}_0 = 3 \sum_{l=2}^{\infty}  (-1)^l{l \choose l-2}{a}_l  . }$$

On the other hand, making no further assumptions on the coefficients of
$ H_2 $, we can isolate two of them in terms of the rest. For simplicity,
we shall isolate $ a_2 $ and $ a_3 $. The result is
$$\displaylines{
a_2 = -{10 m_1 \over R } +{\le \over 3}R^2 + \sum_{l=4}^{\infty} (-1)^l(l-3)
a_l, \cr
a_3 = -{6 m_1 \over R } + \sum_{l=4}^{\infty} (-1)^l(l-2) a_l .} $$
With this in hand we can write the previous expression for the central
value
in terms of $ a_l $, $ l \ge 4 $,
\beq G_{11} (0) = G_{22} (0) = - {3 \over R^2} \Biggl[ {8 m_1 \over R} +
{\le \over 3} R^2 + \sum_{l=4}^{\infty} (-1)^l {(l-3)(l-2) \over 2} a_l
\Biggr]. \eeq

The same can be done for the expression of $ H_2 $ in powers of
$ \tr $, $ H_2 = \sum_{l=2}^{\infty} b_n {\tr}^r $. We obtain
\beq G_{11} (0) = G_{22} (0) = - {3 \over R^2} \Biggl[ {8 m_1 \over R} +
{\le \over 3} R^2 + \sum_{l=4}^{\infty} (l-3) b_l \Biggr], \eeq
where we have used the junction conditions, i.e.
\beq \sum_{l=2}^{\infty} b_n = H_1(1), \quad \sum_{l=2}^{\infty} n b_n =
H'_1(1). \eeq
\subsection{Examples}
\label{ss-e}

We will consider four examples. The first two constitute the
well-known proposals of \cite{ip,fmm,bp} and \cite{dymni,ds}. The
other two constitute a family of new candidates that naturally
arise from the preceding expressions. We start with the second
pair
\subsubsection{Two arbitrary powers}
\label{sss-tap} We choose that only two specific powers of $ H(r)
$, say $ M$, $ N $, be present in this case. In order to fulfill
the regularity conditions, both must satisfy $ M $, $ N \ge 2 $.
However, if it holds that $ M $, $ N > 2$, the Einstein tensor, $
{ G }$, becomes zero at the origin. Thus, if we wish a de
Sitter-like behavior at, and near, the origin, we must impose one
of them to be equal to 2. We then have $ M=2 $, $ N \ge 3 $ as the
only suitable choice for a two power case. $ H_2(\tr) $ reads \beq
H_2(\tr) = b_2 r^2 + b_N r^N, \quad N \ge 3, \eeq with \beq b_2 =
{2m_1 \over R}\Biggl({N+1 \over N-2} \Biggr) + {\le R^2 \over 3},
\\ b_N = -{6m_1 \over R}{1 \over (N-2)}, \quad N \ge 3. \eeq $
G_{11}(\tr) $ and $ G_{22}(\tr) $ read \beq G_{11}(\tr) = -\le +
{6m_1 \over R^3} \Biggl({N+1 \over N-2} \Biggr) \bigl(\tr^{N-2} -
1 \bigr), \\ G_{22}(\tr) = -\le + {6m_1 \over R^3} \Biggl({N+1
\over N-2} \Biggr) \Biggl({N \over 2}\tr^{N-2} - 1 \Biggr). \eeq
Whence one readily sees that their finite value at the origin
coincides, as expected, \beq G_{11} = G_{22} = -\le - {6m_1 \over
R^3} \Biggl({N+1 \over N-2} \Biggr), \eeq and it does not change
significantly if the power $ N $ changes.
\subsubsection{Lowest powers}
\label{sss-lp}
This example chooses the case in which $ H_2 $ has the minimal
power dependence. This amounts to taking $ N= 3 $ in the preceding
example or setting $ a_l = 0 $, $ l \ge 4 $, in the general
expressions. Its interest lies in considering the simplest
situation. The result is \beq H_2(\tr) = \Biggl({8 m_1 \over R} +
{\le R^2 \over 3} \Biggr) \tr^2 - {6 m_1 \over R} \tr^3, \eeq and,
near the matching hypersurface, \beq H_2 (\tr -1) = \Biggl({2m_1
\over R} + {\le R^2 \over 3} \Biggr) + \Biggl({-2m_1 \over R} + {2
\le R^2 \over 3} \Biggr) \bigl(\tr -1 \bigr) \cr
+ \Biggl({-10m_1 \over R} + {\le R^2 \over 3} \Biggr) \bigl(\tr -1 \bigr)^2 
- {6m_1 \over R} \bigl(\tr -1\bigr)^3.
\eeq And \beq G_{11} = -\Biggl(\le + {24 m_1 \over r^3} \Biggr) +
{24 m_1 \over R^3} \tr = - \le + {24 m_1 \over R^3}\bigl(\tr - 1
\bigr), \\ G_{22} = -\Biggl(\le + {24 m_1 \over r^3} \Biggr) + {36
m_1 \over R^3} \tr =  {12 m_1 \over R^3} - \le + {36 m_1 \over
R^3}\bigl(\tr - 1 \bigr) . \eeq $ G_{11} $ tends to $ - \le $ as $
\tr $ tends to 1, in accordance with Israel's conditions.
\subsubsection{The approach of Israel and Poisson}
\label{sss-aip}
In reference \cite{ip} a plausible candidate for the energy-matter
content of the interiors of regular black holes was proposed.
Because of the impossibility of matching directly the de Sitter
spacetime with the Schwarzschild one, the authors proposed that a
singular layer of non-inflation\-ary material should exist between the
de Sitter core and the external Schwarzschild metric. The authors
recognized this assumption as an attempt at overcoming the
problems above mentioned. However the usual spirit of matching a
stellar interior with a vacuum exterior was lost, the reason being
the unavoidable presence of a singular layer, which should be placed
close to the matching hypersurface and would act as a matter
surface density. The authors themselves already pointed out this
problem (see \cite{bp}). They considered that their approach could
be improved by avoiding such a layer and by imposing a smooth
transition from the hypersurface to the de Sitter core. In any
case it was the only available candidate to continue the studies
of quantum regular black holes. Further work, by these authors and
other groups that were dealing with the same problem, addressed
this idea as well as its physical consequences and properties (see e.g.
\cite{markov,fmm,bp}).

The task here will be to see whether this geometrical and physical
model can be recovered from our analysis, {\it solely}
based on the usual spirit of collapsing bodies, i.e. without
involving {\it any} (singular) material layer. First of all,
let us notice that, in our models, $ G_{00}$ can be re-expressed
as \beq G_{00} = {1 \over r^2} (H r)', \eeq where $ ()' $ is the
ordinary derivative with respect to $ r $.

We now search for a solution within our family that tends to that
particular solution (see, for instance, \cite{ip,fmm,markov1,bp}).
This amounts to saying that we look for a de
Sitter core for small values of $ \tr $ and a quantum contribution
of the type of {\em the square of the characteristic curvature of
Schwarzschild spacetime} near the matching hypersurface, i.e. $
(G_{00})_{\rm ext.} \allowbreak \propto m_1^2/r^6 $.
This is different
to assuming that in the exterior region, close to the matching
hypersurface, the quantum contributions turn into a
cosmological-like term. We thus have a quantum exterior different
from the one encountered in the rest of the examples. Obviously, a
combined model could also be solved. However we prefer to maintain
the spirit of the referred works for easier comparison of our
results with theirs. In fact, at the end of Sect.~\ref{ss-nr}, we
will show that the exterior vacuum contribution plays a secondary
role, regardless of its precise form. All these features taken
into account, we first put, generically, \beq G_{00} = {1 \over (B
+ C r^3)^2}, \eeq where $ B $ and $ C $ are two constants, to be
determined after imposing the matching conditions as well as the
mentioned physical requirements. The matching conditions lead to
$$\displaylines{ {1 \over B(B+Cr^3) }= {3 \over R^2} H_1(R), \cr
{2B- C R^3 \over B (B+ CR^3)^2} = {3 \over R}H_1'(R),} $$ where $
H_1 $ comes, as usual, from the external model.

In order to reproduce the new exterior vacuum model, we see that
it is possible to select it from our family, by setting
$ B_{\rm ext.} = 0 $, and $ C_{\rm ext.}^{-1} = \alpha m_1 $,
where $ \alpha = \beta L_{\rm Pl} $, being $ \beta $ of order
unity, and $ L_{\rm Pl} $ the Planck length ($\alpha$ is of order
unity in Planckian units).  $ \beta^2 $ is related to the number
and types of the quantized fields, \cite{ip,bp}. This choice
yields \beq H_1(r) = {2 m_1 \over r}- {1 \over 3}\biggl( {\alpha
m_1 \over  r^2}\biggr)^2, \eeq where we have taken into account
that the exterior region is dominated by the Schwarz\-schild
geometry, for large values of $ \tr $.

Finally, $ B $ and $ C $  yield
\beq
B = {\alpha \over 6- \displaystyle{\alpha^2 m_1 \over R^3}}, \quad
C = {2 \over \alpha m_1}\left(1-{ 3  \over 6 -
\displaystyle{\alpha^2 m_1 \over R^3} }\right).
\eeq

Thus, we do obtain a model within our family that copes with the
desired requirements of references \cite{ip,fmm,bp}. In order to
check if the solution obtained coincides in fact with that
particular model, one has to solve the problem of getting $ R $
from observed or expected values of the constants involved. This
will be postponed to the following section. There, the
plausibility of all these models will be analyzed in detail. To
summarize, we have proven here that a --everywhere smooth---
spacetime model in our
family satisfies all the required geometrical assumptions as well
as the particular form of $ G_{00} $ corresponding to the above
mentioned references.

\subsubsection{Dymnikova's model}
\label{sss-dm}
Some time after the appearance of the previous cases a new
model for a regular
interior of a black hole was proposed, see \cite{dymni,ds}.
However the
approach was now different to that of the previous authors.
In order to avoid the
problem of the singular layer and to introduce a smooth transition
between the
Schwarzschild dominated region and the de Sitter central core,
the imposition of a
clear matching hypersurface was relaxed. Even more, these
authors
did not impose any matching between the two different
spacetimes at
all. The problem was then that the final picture of the
collapsed object departed again from the usual, expected one. Now
Schwarzschild was only recovered in an asymptotical
sense, for $ \tr $ approaching
infinity only.\footnote{A similar assumption is made in \cite{ayon} for the 
case
of {\em charged}, non-rotating black holes. For the time being, here we will 
only
consider those cases of major astrophysical interest, i.e. non-charged black 
holes.
However, an analogous study is straightforward for their model (the core 
however is no longer described by the de Sitter spacetime or som well-known 
spacetime). A distinguished property of that model will be considered in 
Sect.~\ref{s-ecs}.} This immediately adds a natural
drawback in the model, since
collapsed objects have their matter content
restricted to a finite volume. However, if
a sufficiently quick convergent matter model
can be obtained, then the lost mass,
for a observer outside the horizon of the collapsed body
could become as
negligible as desired. Thus one would, at least,
recover a trial model,
interesting enough to support or reject the
conclusions of the previous authors. In this
sense Dymnikova's model is quite interesting.
In a later work, together with B.
Soltysek, they incorporated the observational
fact in favor of a non-vanishing cosmological
term in the exterior region \cite{ds}. This work
includes the first pure Schwarzschild exterior
case also straightforwardly. We shall deal in this subsection
with that model, considering an exterior $ \l $,
or $ \le $ and a definite end of the collapsed body.

The imposition for the energy-matter content is of the form \beq
G^0_0 = G^1_1 = \beta \exp{(-\alpha \tr^3)} + \gamma, \eeq where
the constants $ \alpha $, $ \beta $, and $ \gamma $ are found
after imposing our matching conditions. First of all we can
already impose that the exterior geometry is of the
Schwarzschild-de Sitter type, setting $ \lim_{\tr \to 1} G_{11} =
- \le $. This yields \beq \gamma = -\le - \beta \exp{(-\alpha)} .
\eeq We then integrate the expression of $ G_{11} $ (this is
straightforward because of the assumed $ r^3 $ dependence in this model as 
it was  also
the case in the previous model) in
order to obtain $ H_2 $, getting \beq H_2(\tr) = {R^2 \over 3}
\biggl\{ \biggl({\beta \over \alpha}\biggr)\biggl[{\exp{(-\alpha
\tr^3)} - 1 \over \tr}\biggr] + \tr^2 [\beta \exp{(-\alpha)} +
\le] \biggr\}, \eeq where we have already imposed the regularity
conditions at the origin. The final step is to impose the matching
conditions at the spatial hypersurface. They yield the condition
\beq \beta \biggl[{\exp(-\alpha) -1\over \alpha} + \exp(-\alpha)
\biggr] = {6m_1\over R^3}. \eeq Near the origin the expression of
$ G_{11} $ tends to $ \beta[1-\exp(-\alpha) ] - \le $. This
constant characterizes the geometry of the core. Let us now assume
that it is of de Sitter type as in the model of \cite{ds} (see
also next section). Therefore, it has to be equal to an internal $
\l $, say $ \li $, as opposed to $ \le $, which accounts for the
exterior de Sitter part. Combining both equations, one gets \beq
{\exp(\alpha)-\alpha-1 \over \alpha[\exp(\alpha)-1]}
= {6m_1 \over R^3(\li-\le)}. \eeq This equation is transcendental.
In order to solve it one realizes that there always exists a value
of $ R $, big enough, in order to make $ 6m_1/R^3(\li-\le) $ small
enough. The reliability of this restriction of ``big enough $ R
$'' will be considered later. The numerical results of the next
section will show that the associated error is indeed very small.
Our aim is now to see whether the model considered by
\cite{dymni,ds} lies also within our family of models.

Within the restriction above, one can see that the solution is
unique and appropriately given by
\beq
\alpha = {(\li - \le) R^3 \over 6 m_1}, \quad \beta = \le - \li .
\eeq
The relative error being
$ \Delta = \alpha/[\exp(\alpha) - 1] \sim \alpha\exp(-\alpha) $.
Therefore, it can be
made as small as needed, provided $ R $ can be taken as
large as necessary. The
coefficient $ \alpha $ can be used to define a characteristic
radius, say $R_c $, as
\beq
R_c^3 \equiv {R^3 \over \alpha} = {6m_1 \over \li - \le}
= {3 R_g \over \li - \le},
\eeq
where $ R_g $ is the Schwarzschild radius of the object. With
this, for $ 0 \leq r \leq R $,
$ G_{11} $ and $ G_{22} (= - \le + \beta \exp(-\alpha \tr^3)
(1-3\alpha\tr^3/2) $) can be rewritten as
\beq
G_{11} = - \le + (\le - \li ) \exp[-(r/R_c)^3], \\
G_{22} = -\le + (\le-\li)\biggl[ 1- {3 \over 2}
\biggl({r \over R_c}\biggr)^3 \biggr] \exp[-(r/R_c)^3],
\eeq
which exactly coincide with the expressions in the mentioned papers
(see e.g. Eq.~(11), (14) of \cite{ds}), although they have now been obtained 
for the physical
range $ 0 \leq r \leq R $.
\section{Significance and interpretation of the conditions obtained}
\label{s-sico}
\subsection{Physical consequences of isotropization}
\label{ss-pci}
Until now the precise value of the matching $ r $ (or $ R $) has
been overlooked. In part, because it was actually not necessary to
fix it, that is, the conditions hold at any value of $ r $, which
is actually a rather remarkable fact. Despite the interesting
observation that all the previous properties are independent of
the precise value of $ R $, the final aim is to match the
preceding results with physical observations and with well-known
theoretical behaviors of the quantum state near the origin, as
well.

Isotropization is a general property, as we have pointed out, and
it can be interpreted as a geometrical one. By this we mean that
it is actually the Einstein tensor that becomes isotropic,
regardless of the spacetime chosen, among our general family. If
we write the Einstein tensor in terms of the energy-momentum
tensor and the $ \l $ contribution, we realize that, at the
origin, the following relation must hold \beq { G}(0) = {8 \pi G_N
\over c^2} { T }(0) - \li { g}(0). \eeq That is, both the vacuum
contribution and the rest of the contributions of the excited
states of the quantum field add directly. However, we can break
this indeterminacy with the help of some physical model of quantum
behavior near the origin. According to current ideas, we expect
that the contribution from the vacuum will dominate those of the
rest of states. Thus we simply impose $ { T} (0) $ to be
negligible or zero when compared with $ \rho_{\li} $($ =(c^2/8\pi
G_N)\li $). On the other side, $ { G} (0) $ is in all cases a
value that depends on $ R $, $ m_1 $, and $ \le $ only. Therefore,
we are faced up with a general relation of the type \beq
{ G}(m_1, \le; R) = -\li { g}
\eeq
or
\beq
G_{11}(m_1, \le; R) = -\li.
\eeq
This relation will allow us to find $ R $ in terms of $ m_1 $ and
$ \li - \le $ only. The expressions of $ G_{11} $ in more general
situations have been given before.

Finally, let us add that the model chosen here can be
reinterpreted as having imposed an isotropization of the quantum
gravitational effects in favor of the quantum vacuum. It is a new
isotropization added to the geometrical one, which in our family
has a universal character.
\subsection{Numerical results}
\label{ss-nr} According to several recent observations
\cite{perl,ries}, in what follows we shall assume that $
\rho_{\le}\! \in \! [10^{-10}, 2 \times 10^{-8}] \;{\rm erg
\!\cdot\! cm^{-3}} $ . However an analysis shows that essentially
the same results will hold even if $ \rho_{\le} $ is {\it zero},
or another type of quantum vacuum contribution is assumed,
as was the case in the Israel-Poisson model. The fundamental contribution 
comes
directly from the quantum gravitational model that is imposed for
the core. For the reasons given before, we shall consider, for the
time being, an internal de Sitter-like core. Nonetheless, there is
no present agreement about the scale at which regularization could
act. This is directly connected with the so called ``{\em
Cosmological Problem}'', which is the subject of intense research
currently.\footnote{The possibility of dealing with two $ \l $'s
has not been stressed yet and will not be discussed here.
Elsewhere, we will try to perform a detailed theoretical analysis
on the plausibility of having $ \le - \li $ of the order of the
used numerical ($ \le $) and observational ($ \li $) values.} Our
strategy will be to let this constant free and then consider the
different results that will come out. A convenient way to handle
and integrate this indeterminacy is to set $ \li = 10^{3s}\le $,
being $ 3s $ the free parameter that governs the scale of
renormalization. For instance, if $ s $ is around 40, we are then
considering that regularization takes place at Planck scales, etc.
\subsubsection{Two arbitrary powers}
\label{sss-tap2}
The fundamental relation is
\beq
\le = \li + {6m_1 \over R^3}
\Biggl({N + 1 \over N - 2} \Biggr), \quad \forall N \ge 3,
\eeq
whence
\beq
R = R_{\odot} \root{3}\of{M} \root{3}\of{N+1 \over 4(N-2)},
\eeq where we have put $ m_1 = M m_{\odot} $, $ m_{\odot} $ being
the Sun mass, and $ R_{\odot} \equiv $ $ \root{3}\of{24m_{\odot}/
}\allowbreak\overline{(\li -\le) }$. This value only depends on
the regularization scale and, in fact, is in correspondence with
the solution for a collapsed object of one solar mass in the case
of the ``lowest powers'' model: $ R_{\odot} \in [3 \times
10^{21-s}, 6 \times 10^{20-s}] {\rm cm} $. For $ s= 40 $ we get $
R_{\odot} \in [3 \times 10^{-19}, 2 \times 10^{-20}] {\rm cm} $.
Yet we see that the object has a quantum size very far from
Planckian scales even if $ s $ is bigger. In general $ R_{\odot} /
L_{\rm Pl} \ge 10^{13}!$ Moreover this result is valid for all $ N
$ since for any $ N $ we have that $ R \in [0.6  , 1] R_{\odot}
\root{3}\of{M} $. It is obvious that for any astrophysical object
the final properties are very similar. Table~\ref{tab-1} comprises
different massive objects and regularization scales.
\begin{table}
$$
\begin{array}{||c||c|c|c|}
\hline\hline
m_1 & s= 30 & s= 40 & s= 50 \cr
\hline
\hline
M_{\odot}  & 10^{-9}& 10^{-19} & 10^{-29} \cr
\hline
10^3 \,  M_{\odot} & 10^{-8}& 10^{-18} & 10^{-28} \cr
\hline
10^6 \, M_{\odot} & 10^{-7}& 10^{-17} & 10^{-27} \cr
\hline
10^9 \, M_{\odot} & 10^{-6}& 10^{-16} & 10^{-26} \cr
\hline
\end{array} $$
\caption{\label{tab-1} {\small $ R $ in cm for various
astrophysical and galactic objects and
different scales of regularization ($ s = 30 $ corresponds
to a GUT's regularization scale, $ s = 40 $ to a
Planckian one, etc.). In any
case $ R/L_{\rm Reg} $ is much bigger
than 1 ($ R/L_{\rm Reg}
\sim 10^{(-6 + s/2)}$). Therefore all
of them are mainly in the semiclassical regime.}}
\end{table}
\subsubsection{Israel-Poisson's approach.}
\label{sss-ipa}
We have found that the corresponding model within our family
must satisfy
\beq
B = {\alpha \over 6- \displaystyle{\alpha^2 m_1 \over R^3}}.
\eeq
In this case, $ B^{-2} = \lim_{\tr \to 0} G_{00} = \li $, so that
\beq
R^3 = {\alpha^2 m_1 \over 6 - \alpha\sqrt{\li}}={\beta^2 \over 6
- \beta \sqrt{\li L_{Pl}^2}} m_1 L_{\rm Pl}^2 . \eeq This model
clearly depends on the coefficient $ \beta $. For instance, in
order to obtain a solution, we must have $  \beta^2 < 36/(\li
L_{\rm Pl}^2) $. The natural scale of regularization in this model
is the Planckian one since from the beginning the coefficient $
\alpha $ was related to the Planck length. Obviously other
regularization scales would simply change $ L_{\rm Pl} $ by the
corresponding scale. Using standard values for $ \li $ that use
a Planckian regularization scale, and that $ \beta^2 $ should
be at most of order unity \cite{ip,bp}, we get $ R \sim
\root{3}\of{M} \times 10^{-20} {\rm cm.} $ This result is in
complete agreement with the foregoing values, even though the
models are very different, from a physical point of view.
\subsubsection{Dymnikova's model}
\label{sss-dm2} In the original model there is no frontier (hypersurface)
up to which the spacetime is of exact Schwarzschild-de Sitter
type. Thus a radius cannot be defined at which the collapsed
object ends.
However, a characteristic radius, $ R_c $, is found there, and
  it is proven that, for $ (r/R_c )^3 >> 1 $, the outside matter can
be dismissed in comparison with the inner one. This was the main
reason  in order to consider that such model describes
approximately an object of quantum size.

In building an analogue of that situation, we decided to introduce
three parameters, namely $ \alpha $, $ \beta $, and $ \gamma $ in
order to better mimic the model for $ G_{11} $. In the rest of the
models there are only two parameters. Now we have an extra
freedom. In fact, for any $ R > \root{3}\of{2} R_c $, there exists
a unique solution for them, provided $ \le $ is less than $\li $,
as we expect. For the rest of the values, there is no solution at
all. Thus we are faced with an expected freedom in $ R $ that can
be dismissed adding a further assumption. For the time being let
us see that there are solutions which fit in a very precise way
with the previous results.

This model was solved explicitly for the case in which $ \alpha $ was big 
enough so
that the relative error, $ \alpha\exp(-\alpha) $, was
small. Let us redefine $ R $ as $ R= k \root{3}\of{M} R_{\odot} $,
where $ k $ is an adimensional constant, to be found below. Thus
\beq
\alpha = \Biggl({R \over R_c}\Biggr)^3 = 4 k^3.
\eeq The relative error is then $ \Delta = 4k^3\exp(-4k^3) $.
Whence, we see that choosing $ k $ of order unity the relative
error is very small (for instance, for $ k=2 $ it is only of $
10^{-11} \, \% $!). Thus $ R= k \root{3}\of{M} R_{\odot} $, with $
k $ of order one, are all very good solutions of our corresponding
model. Since there is no well
defined frontier for the collapsed object
in the approach of \cite{dymni,ds},
a well-defined $ R $ does not
exist. There, the assumption
about the characteristic size of the object was ascribed to $ R_c
$. In fact, $ R_c = R/(k\root{3}\of{4}) $, so that the results in
\cite{dymni,ds} fit well with ours.

Finally, we may add a further restriction in our model in order to
eliminate the freedom in $ R $. Before we proceed, it is worth
remarking that the above discussion was just a check of the
issue, if a model of the type \cite{dymni,ds}
was inside our family. In fact, this is only a completion criteria
that allows us to continue dealing with this model. It is commonly
thought that no known quantum fields can be responsible for such a
$ G_{11} $.

A natural way to put an end to the indeterminacy in $ R $ is by
setting, e.g.,  $ \gamma = 0 $. With this assumption we have chosen an
option that completely separates the quantum behavior in both
regions. Obviously, many other choices are possible. The result of
such choice is \beq
R^3 = {6m_1 \ln(\li/\le) \over \li - \le - \le \ln(\le/\li)}.
\eeq Now, taking into account that $ \li = 10^{3s} \li $, and
recalling that we expect $ s \ge 40 $, we obtain
\beq
R \sim 2 R_c \root{3}\of{s}.
\eeq For $ s=40 $ this yields $ R \sim 7 R_c $, which again falls
within the same order as all the previous models. This fact leads
us to believe that the main properties found hitherto are rather
general and independent of the model that describes the collapsed
object. Notice finally that the initial considerations that came
from a direct, and wrong, matching between a de Sitter core and a
Schwarzschild exterior yielded $ R = R_c $. So we can assert that
the correct solution modifies the naive, and wrong, expected value
by a small factor. This is quite an amazing conclusion.
\section{Horizons and an interpretation of the regularized black hole}
\label{s-hirbh}
Looking, for instance, at Eq. (30) in \cite{ds} and
comparing it with our result
\beq
g_{00} = -1 + H_2 ,
\eeq we realize that substituting here our corresponding $ H_2 $
for their model, these expressions turn out to be very similar
same, except for an overall sign due to the different signatures (
(+,--,--,--) instead of our (--,+,+,+)). We conclude that the same
structure for the horizons and Cauchy hypersurfaces is obtained.
In \cite{ds} the solutions are obtained by approximations of the
exact solution, so that these results and ours are really
coincident (the relative error being completely negligible with
respect to both exact solutions).

In general, the horizons result from the cancellation of $ g_{00}
$. Thus we are left with a general set of horizons. A global
study, for all the candidates encountered, has not been carried
out yet. We could focus on examples, and try then to extract some
general features from them, but we do not find this of primary
importance.\footnote{With respect to the other models found here,
we have got that the results are rather similar to those
of Dymnikova and Soltysek's model
\cite{ds}.} The main point is here, in fact, that the matching
occurs at a radius which is substantially smaller than the
Schwarzschild radius of the object. Therefore we will always have
a typical exterior, a vacuum transition region extending until the
matching with the object happens, and a quantum-dominated
interior, which finally converges to a de Sitter core. In the
vacuum interior region and in some part of the quantum object, the
role of $ t $ and $ r $ are not interpreted as usual ($\partial_t
$ changes its character). This is the reason for adequately
treating the horizons: to see where exactly such changes appear.
But, we can still perfectly agree in ordinary physical terms
without requiring a general resolution of the precise radii at
which horizons occur.

Another issue is that of the topology of the solutions, if they
are regular black-holes, and their possible extensions, i.e.
``universe reborn''. Its general structure can be found in
\cite{borde}, where it was shown that the topology of any regular
black hole should be similar to that of a singular-free
Reissner-Nordstr\"om spacetime. Thus, there appears a necessary
topology change, if one deals with their complete extension.
\section{Energy conditions}
\label{s-ecs} A common point in dealing with the avoidance of
singularities is to show that the required energy conditions in
the singularity theorems (see e.g. \cite{he}) fail to be valid.

Here we will study three local energy conditions, commonly
considered in the literature: strong energy conditions (SEC), weak
energy conditions (WEC) and null energy conditions (NEC). SEC are
related with the formation of singularities in the collapse of an
object. WEC are directly related with the energy density measured
by an observer and NEC are useful in order to include some
spacetimes which violate the first two, but are predicted by some
quantum models, e.g. anti de Sitter spacetime. Although an
analysis of energy conditions helps to understand the physics of a
model, one has to be cautious ascribing to them more relevance
than they actually have. In several systems, mainly when quantum
effects play a fundamental role, they all may be violated with
less difficulty (see e.g. the review in \cite{visser}). Our
results also agree with those in \cite{mps}, where the authors
study them in general spherically symmetric spacetimes.

Let $ \{ {\vec e}_a \} $, $a=0,1,2,3$, be a dual vector basis of the cobasis 
in~\rf{cobasis},
defined by
$ \th^b {\vec e}_a = \delta^b_a $, $ b=0,1,2,3$. Any timelike vector field,
$ {\vec V} $, in this manifold can be represented by
\beq
\label{4vel}
{\vec V} = A^b {\vec e}_b, \qquad (A^0)^2 = 1 + \sum_{i=1}^{3} (A^i)^2,
\eeq
where $ A^b $ are some functions.

On the other hand, from the results of Sect.~\ref{s-fsus}, the
Ricci tensor is \beq {\bf Ricci} = R_{00} (\th^0\otimes\th^0 -
\th^1 \otimes \th^1) + R_{22}(\th^2\otimes\th^2 + \th^3 \otimes
\th^3), \eeq where $ \otimes $ is the tensor product. A similar
expression is valid for the Einstein tensor.

SEC require $ R_{VV} \equiv R_{ab}V^aV^b \ge 0 $, for all $ \vec V $. From 
the
expressions above, we obtain $ R_{VV} = R_{00} + (R_{00} +
R_{22})[(A^2)^2+(A^3)^2] $. Taking into account Eqs.~\rf{ricci}
and~\rf{einstein}, $ R_{VV} = G_{22} + (G_{00} +
G_{22})[(A^2)^2+(A^3)^2] $. Finally, using Einstein's
equations,~\rf{Einstein}, and the fact that $ A^2 $, $ A^3$ are
free, we get \beq
{\rm SEC} \leftrightarrow \rho + p_2 \ge 0, \quad p_2 \ge 0
\eeq
where $ \rho $ is the energy density measured by $ {\vec e}_0 $,
$ 8 \pi \rho = G_{00} $ and $ p_2 $ is the tangential pressure (or stress) 
of
the source, $ 8 \pi p_2 = G_{22} $.
This is the usual representation of SEC. However, the GNRSS
family allows for a different, more useful, expression. Indeed, as
mentioned elsewhere, it is easy to show that, for a GNRSS
spacetime, $ p_2 = -(\rho + r\rho'/2) $,  where $ ()' \equiv d()/dr $.
Therefore, we can
eventually write \beq
{\rm SEC} \leftrightarrow p_2 \ge 0, \quad \rho' \leq 0 .
\eeq
Following analogous steps, one finds, for WEC ($G_{VV} \geq 0 $, for all $ 
\vec V $)
\beq
{\rm WEC} \leftrightarrow \rho \ge 0, \quad \rho' \leq 0 .
\eeq
In the case of NEC, $ \vec V $ is a {\em null} vector field, $ \vec V \cdot
\vec V = 0 $, and requires
the evaluation of $ R_{ab} V^a V^b = G_{ab} V^aV^b \ge 0, \forall \vec V $. 
One
obtains
\beq
{\rm NEC} \leftrightarrow \rho' \leq 0 .
\eeq
Thus one sees, that a necessary condition {\em common to all
of them} is that the energy profile of the sources be a
non-increasing function. Moreover, in WEC, $ \rho \ge 0 $ conveys
the positivity of energy density whereas, in SEC, $ p_2 \ge 0 $
refers to the positivity of tangential pressures. The last is in
fact violated from some value of $ r $ downwards, as can be
readily seen from $ p_2 = -(\rho + r\rho'/2) $ and the fact that
the source is regular at the origin. Therefore, eventually, the
singularity is not created.
\subsection{Energy conditions for different models}
It is now easy to evaluate the fulfillment or violation of the
energy conditions in the models presented before. It turns out
that WEC, and NEC, are satisfied in all of them very easily for
any value of $ r $ (e.g. for (anti) de Sitter core, $ \rho'=0 $).
One only needs to impose $ \le > \li $. On the other hand, as
mentioned before, SEC are violated in all of them. The new result
is that this occurs far away from the regularization scale.
Indeed, a detailed analysis shows that SEC are violated for $ r
\leq R_{SEC} $, with \beq R_{SEC} = \root {N-2}\of{2 \over N} R,
\quad R_{SEC} = \left( \root{3}\of{ \alpha^2 m_1 \over
4(6R^3-\alpha^2m_1)} \right) R, \quad R_{SEC} = \root{3}\of{2
\over 3} R_c, \eeq where, all the quantities have been defined in
Sect.~\ref{ss-e} and the solutions correspond to the two-power
model, the Israel-Poison's model and  Dymnikova's model,
respectively. It is clear from these results, that SEC are
violated in the most part of the object, i.e. $ R_{SEC}
\stackrel{<}{\sim} R $ (see also \cite{dymni}). For the evaluation
of the Israel-Poison's model, we have used the same numerical
values as in Sect.~\ref{ss-nr}.

Let us now consider the series of models in \cite{ayon}--\cite{ayon3}.
In \cite{ayon} the first regular models for non-rotating black holes were 
given.
We remark that all these solutions are
{\em charged} black holes,
e.g. $ |e|/m $ is of order unity, where $ e $ is the charge
of the black hole. This is far away from astrophysical
observations of black holes (see e.g. \cite{astro}). Nevertheless,
these solutions are very important in this issue, since they are
the {\em first} solutions to regular black holes with a clear
interpretation in terms of a known quantum field, in this
case nonlinear electrodynamics. A direct analysis
of those works shows that, unexpectedly,
{\em all} of them belong to the GNRSS family of
spacetimes. Let us choose one of the three solutions proposed.

For instance, the one in \cite{ayon3} sets, in our notation,
$$ H = {2m \over r} \biggl[ 1 -\tanh\biggl({e^2 \over 2mr}\biggr)\biggr].
$$ The energy density can be computed from Eqs. \rf{einstein}. We
obtain $$ 8 \pi \rho_{AG} = { e^2 \over r^4 \cosh^2(e^2/2mr)} = {1
\over e^2 s^4} {1 \over y^4 \cosh^2y^{-1}}, $$ where, in the last
expression we have used the same notation as the authors, that is,
$ s \equiv |e|/2m $ and $ y \equiv 2mr / e^2 $. It is easy to see
that $ \rho_{AG} $ is always positive. However, $ \rho_{AG}' $
becomes positive from some $ y_0 $ downwards. In fact, a direct
computation shows that the maximum energy density is achieved for
$ y_0^{-1} \tanh(y_0^{-1}) = 2 $, i.e. $ y \sim .4842 $, and that $ \rho $
is an increasing function for $ y \leq y_0 $. Finally, taking into
account \cite{ayon3} that $ |e| $ must be lower than $ 1.05 m $,
in order to have a regularized solution, we can readily show that
for any $ r \leq 0.2664 m $, that is, the {\em most} part of the
object, {\em all} energy conditions fail, even WEC and NEC. The
rest of proposed models behave in an identical way.\footnote{In
\cite{ayon} the authors even give a graphical representation of $
-g_{tt} $,
whence direct visualization proves our assertions.} This adds
a new (elementary) example
to the violation of energy conditions models when quantum effects
play an important role (see \cite{visser} for a recent review)
and shows us that energy conditions help
understanding the models, but not necessarily should bound the
search for new solutions.

\section{Quantum aspects of the models}
\label{s-qam} To find the sources for  the models presented is not
easy, but this will be a  necessary step to undertake in order to
show the plausibility for the avoidance of the singularities. To
reach this goal  within our scheme, one has in fact to quantize
the sources. The symmetries of the matter-energy tensor, i.e. $
\rho + p = 0 $, where $ p $ is the radial pressure (or stress)
---$8 \pi p = G_{11} $--- and the one coming from spherical symmetry, 
$ p_2
=p_3 $ are fundamental in order to probe which type of quantum
fields could correspond to their sources. Taking into account the
spherical symmetry of the spacetime and, consequently, of the
Einstein tensor, one easily checks that all classical fields
adapted to the spherical symmetry satisfy this condition (see for
instance Sect. 3.8 of \cite{bd}). This bonus was indeed expected,
because spherical symmetry is a basic geometric symmetry in the
scheme. On the other hand, the other Eq., $ \rho + p = 0 $, can be
rewritten in a general covariant way, without referring to any
special set of observers. Its expression is then \beq T_{\ell
\ell} \equiv T_{\mu \nu}l^{\mu} l^{\nu} = 0 , \eeq
where $ \vec \ell $ is the geodesic radial null direction characteristic
of the GRNSS spaces. If one considers a scalar field, one has
\beq
\begin{array}{rcl}
T_{\ab} & = & ( 1 - 2 \xi) (\nabla_{\alpha} \phi) (\nabla_{\beta} \phi)
+ (2 \xi - {\textstyle{1 \over 2}}) g_{\ab}(\nabla^{\rho}\phi)
(\nabla_{\rho}\phi) \cr
& & - 2 \xi \phi \nabla_{\alpha} \nabla_{\beta} \phi
+ {\textstyle {2 \over n}}\xi g_{\ab} \phi \dal \phi
- \xi \Bigl[ G_{\ab} + {\textstyle{2(n-1) \over n}} \xi
R g_{\ab} \Bigr] \cr
& & + 2 \Bigl[ \textstyle{1 \over 4} - ( 1 -
{\textstyle{1 \over n}}) \xi \Bigr] m^2
g_{\ab} \phi^2,
\end{array}
\eeq
where $ \xi $ is a constant representing the coupling
between the scalar and the gravitational field, $ \phi $ is
the (classical) scalar field, and $ n $ is the dimension of
spacetime. Whence we obtain
\beq
\label{tll}
T_{ll} = (1-2\xi){{\dot \phi}}^2 - 2 \xi\phi{\ddot \phi},
\eeq where $ ()^{\cdot} \equiv \ell^{\lambda} \nabla_{\lambda} $, and
where we have used the fact that $ \lh $ is null and geodesic. If
the field were a classical one, we should have \beq
\label{cond-ll} (1-2\xi){\dot {\phi}}^2 - 2 \xi\phi{\ddot \phi} = 0. \eeq 
The
first thing one notices is that there is no $ \xi $ that makes
Eq.~\rf{cond-ll} be identically satisfied. For instance, in the
conformal case, the trace of $  T $ is the central object. Then,
choosing $ \xi = -(n-2)/4 (n-1) $, the conformal coupling, and $ m
= 0 $, one gets that the trace vanishes for the classical field.
Therefore, Eq.~\rf{cond-ll} imposes now more conditions on $ \phi
$ than in the well-known conformal case. This is a remarkable
difference with respect to conformal field theory. Actually
Eq.~\rf{cond-ll} can be integrated, giving two results depending
on whether $ \xi = 1 / 4 $ or $ \xi \neq 1 / 4 $. In any case,
they must satisfy the field equations.

For the GNRSS family, the free scalar field equations,
$ (\dal + m^2 + \xi R ) \phi = 0 $,
where $ \dal $ is the d'Alembertian operator of each
spacetime, may be written as ($ l = 0, 1, \ldots $)
\beq
\label{eq-field1}
\begin{array}{l}
-(1+H) \partial^2_t \phi + 2 H \partial_{t}\partial_{r} \phi
+ ( 1 - H ) \partial^2_{r} \phi - \partial_r H (\partial_t \phi - \partial_r 
\phi) \cr
\hfill{}+ \Bigl[ m^2 + \xi R - \textstyle{l(l+1) \over r^2} \Bigr] \phi = 0 
,
\end{array}
\eeq where $ R $, the scalar curvature, is given by
\beq R  = {2H \over r^2}+ {4 {\dot H} \over r}
+{\ddot H}, \eeq and where a separation of variables has been
done using the spherical symmetry of the problem. Hence, $ \phi $
in Eq.~\rf{eq-field1} is a function of $ t $ and $ r $ only. We
notice that $ R $ and Eq.~\rf{eq-field1} may be written more
intrinsically as $ R = (2 / \theta^2) {\ddot (H \theta^2)} $, where $
\theta = 1/2r^2 $ is the expansion of $ \lh $, and \beq
\label{eq-field2} 2 \partial_u {\dot \phi} - 2 {\dot H} {\dot \phi}
- 2 H {\ddot \phi} +
\Bigl[ m^2 + \xi R - { l (l+1) \over r^2} \Bigr] \phi = 0, \eeq
where $ u \equiv  (1/\sqrt{2})(r-t) $.
It turns out eventually that the solutions of Eq.~\rf{cond-ll} do
not satisfy the free scalar field equations. Furthermore, the end
result is that no free classical field satisfies Eq.~\rf{cond-ll}.

In order to see whether a quantum treatment could make
Eq.~\rf{cond-ll} be satisfied, we shall begin the process of
quantization of the sources for the GNRSS family (we remark that
actually the steps are valid for {\em any} Kerr-Schild-like spacetime).
First, one needs to calculate $ \delta $. In fact, each spacetime
of the GNRSS family is obtained by changing $ H ( r ) $, and
leaving $ \lh $ fixed. This is actually how the different
interiors are obtained. Obviously, since $ \lh $ is null, we could
also choose that $ \lh $ changes to a multiple of it, but this is
unnecessary choosing a convenient definition of $ H $. Therefore,
$ \delta g_{\mu \nu} $ is simply \beq \delta g_{\mu \nu} = 2 (
\delta H ) \ell_{\mu} \ell_{\nu}, \eeq and, because $ g^{\ab} =
\eta^{\ab} - 2 H \ell^{\alpha} \ell^{\beta} $, we also have $
\delta g^{\ab} = - 2 (\delta H) \ell^{\alpha} \ell^{\beta} $.
Finally, $ g = \det(g_{\ab}) = \det(\eta_{\ab} + 2 H \ell_{\alpha}
\ell_{\beta}) = \det(\eta_{\ab}) = - 1 $.

The following step is to write down the action, $ S[g_{\ab}] $, in
terms of $ S[\eta_{\ab}] $. By definition, see e.g. \S 6.3 in
\cite{bd}, \beq T_{\ab}(x) = {2 \over [-g(x)]^{1 \over 2}} {\delta
S \over \delta g^{\ab}(x)} , \eeq where $ S $ corresponds to the
classical action of matter. Performing a standard calculation, we
get \beq S[g_{\ab}] = S[ \eta_{\ab} ] - \int T_{\ell\ell}(g_{\ab})
\delta H \, d^n x . \eeq Whence, \beq \label{tlls} T_{\ell\ell}
[g(x)] = - {\delta S [ g] \over \delta H }\Bigm|_{H =  0}. \eeq
This result may be straightforwardly generalized to include any
spacetime which is in a Generalized Kerr-Schild (GKS)
correspondence with another one, i.e. $ {\bar g}_{\ab} = g_{\ab} +
2 H \ell_{\alpha} \ell_{\beta}$ (see e.g. \cite{kramer}). Notice that the 
GKS form clearly
takes the aspect of a problem with a spacetime acting as a
background, not necessarily flat spacetime.

In our case, we have $ T_{\ell \ell} = 0 $ (in fact, this is the
case of, e.g., any Kerr-Schild metric with a geodesic $ \lh $.
Therefore, the following results are also valid, for instance, for
Kerr-Newman metrics, or some deviations of them, which play a
similar role as the GNRSS here and have been recently proposed,
\cite{burinskii}). Thus, should any classical field exist, this
would mean that the classical action should remain invariant under
a Kerr-Schild relation. We have already shown that this
possibility is forbidden by Eqs.~\rf{cond-ll} and~\rf{eq-field1} 
or~\rf{eq-field2}.
Nevertheless, expression~\rf{tlls} tells us that $ T_{\ell\ell} $
becomes now the essential quantity in the process of quantization,
in the same footing as $ T^{\lambda}_{\, \lambda} $ in the scalar
case. This result points towards the structure of a quantization
of a GKS spacetime. Its study is, as of now, in progress and will
be a matter of further research. It is clear that a complete
solution of the problem in hand depends on this issue.

The aim is now to test if the interiors can be the result of the
quantization of some classical field. To that end, it is enough to
focus the attention on the scalar field. We will therefore study
the possibility that $ < T_{\ell \ell}> = 0 $. This is a natural
option, due to the non-existence of a triad $ (\phi, \xi, m ) $
satisfying Eq.~\rf{cond-ll} and Eq.~\rf{eq-field2} at the same
time. Of course, this is not always a solution. Namely, given a
spacetime metric, $ g_{\ab} $, its Einstein's tensor, $ G_{\ab} $,
will not very often correspond to a known physical source. It is a
general fact of any classical field equations (e.g. the monopole
solution of classical electromagnetism). The same may happen now,
even after having quantized the field. Then, there still would be
a place for a new theory that explains it, or, eventually, it
should be considered as a non-physical solution. It is interesting
that some of the GNRSS solutions are solutions of a non-linear
generalization of electrodynamics, see \cite{ayon}, which is also
connected with string/M-theory.

Coming back to the point, we have that $ < T_{\ell \ell} > $
should vanish, while $ (T_{\ell \ell})_{\rm Cl} $, the classical
contribution, is different from zero for any spacetime belonging
to the GNRSS family.

Let us compare our present situation with that of conformal field
theory and the de Sitter spacetime. There we have that for certain
values of $ \xi $ and $ m $, say the conformal coupling and $ m =
0 $, the classical field verifies \beq (T^{\lambda}_{\,
\lambda})_{\rm Cl} = 0 . \eeq However, one finds after a standard
calculation ($ n = 4 $, $ \hbar  = 1$), \cite{bd,gmm}, \beq
<T^{\lambda}_{\, \lambda}> = -{a_2 \over 16 \pi^2}, \eeq where $
a_2 = -1/15\alpha^4 $, and $ \alpha $ is the ``radius'' of the de
Sitter spacetime. In that way, the de Sitter geometry, i.e. $
R_{\ab} = \Lambda g_{\ab} $, may be interpreted, from a source
point of view, as the semiclassical effect of the quantum vacuum
associated with the conformally coupled, massless, scalar field.
In fact, there are several other possibilities, for other
couplings and masses, but the effect of the anomaly is most
clearly seen in the conformally coupled and massless case.
However, $ T_{\ell \ell} $ obeys the contrary pattern, i.e. $ (T_{\ell
\ell})_{\rm Cl} \neq 0 $, while we pretend that $ <T_{\ell \ell}>
= 0 $. We shall show below that this behavior actually comes from
an {\em identical} anomaly effect.

For the de Sitter spacetime, one has \beq < T_{\ab}> = {T \over 4}
g_{\ab}, \eeq where $ T \equiv < T^{\lambda}_{\, \lambda} > $.
Therefore, since de Sitter spacetime belongs to the GNRSS family,
$ (T_{\ell \ell})_{\rm Cl} $ is different from zero, yet we have $
<T_{\ell \ell}> = 0 $. And this result has been obtained using the
{\em same} standard formalism as that of the semiclassical
approach to gravity. The reason is to be found in the fact that,
for any local observer, we can write \beq < T_{\ab} > =
\pmatrix{\Lambda & & & \cr & -\Lambda & &\cr & & -\Lambda & \cr
& & & -\Lambda}, \eeq and, therefore, $ < T^{\lambda}_{\, \lambda}
>  = - 4 \Lambda \neq 0 $, while $ < T_{\ell \ell} > = T_{00} +
T_{11} = 0 $. Thus, the same reasoning that leads from $
(T^{\lambda}_{\, \lambda})_{\rm Cl} $ to $ < T^{\lambda}_{\,
\lambda} > $ explains, for the de Sitter spacetime, that $
(T_{\ell \ell})_{\rm Cl} $ is non zero, but $ < T^{\lambda}_{\,
\lambda} > $ vanishes.

Finally, we remark that all these models tend to form a de Sitter
core as $ r/R << 1 $. Thus, the de Sitter spacetime will be a good
approximation for all them, lying still in the semiclassical
regime (see table~\ref{tab-1}). In fact, in the models of
Sect.~\ref{ss-nr}, the deviation is proportional to $ (r/R)^3 $,
and the approximation is good enough, provided that $ r/R << 1 $.
On the other hand, the semiclassical approach makes only sense as
an approximation. Therefore, we do not need to include higher
order corrections, as of now.

Another point of view is that $ T_{\ab} $ may be written as $
T_{\ab}  = (T_{\rm de\; Sitter})_{\ab} + \Delta T_{\ab} $. It
turns out that both $ T_{\rm de \;Sitter} $ and $ \Delta T $
satisfy the four Wald axioms (see \cite{wald}), for a $ T_{\ab} $
to be regularized. Thus, we could interpret that here we are
regularizing the de Sitter's part, as if it were a background,
(i.e. the vacuum contribution above), to which one should further
add the contributions of the ``excited'' states, represented by $
\Delta T_{\ab} $. All this shows, at the very least, that the
general scheme of quantization fits quite well within the GNRSS
family.

We recall that those above are just steps towards the confirmation
of a plausible quantum origin for all (or some) of the solutions
in the GNRSS family. To completely fix the issue, one should
either solve Eq.~\rf{eq-field2}, by finding a particular set of
modes, or using Eq.~\rf{tlls}. Regarding the resolution of
Eq.~\rf{eq-field2}, we note that a further separation of variables
between $ t $ and $ r $ is not possible, since the solution should
be valid for the region $ r < R $, where
the spacetime is non-stationary, so that usual mode
decomposition is useless. Another set must be found, compatible
with Eq.~\rf{cond-ll}, that should be matched with some of the
well-known exterior ones,  coming from the solution of
Schwarzschild or Reissner-Nordstr{\"o}m spacetimes. Complete
fixing of the exact problem seems to be a difficult task.

Let us finally resort to a different way, less fundamental but
seemingly more effective, in order to incorporate possible quantum
corrections into the models. This may serve as a temporary
solution, yet one that has been thoroughly used in the literature
to deal with the effects of quantum fields in gravity. The
interesting fact in our context is that, in this way, they are
easily incorporated into {\it any} classical solution of the GNRSS
family.

Indeed, it is only necessary to compute $ <T_{00}> $ to
obtain $ <T_{\ab}> $. This is because any GNRSS spacetime
satisfies  $ <T_0^0> = <T_1^1> $, $<T_2^2> = <T_3^3> $.
And, additionally,  conservation of $ <T_{\ab} >$ yields $ <T^2_2>
= <T^0_0> + (r/2) <T^0_0>^Ò $ (see also footnote \ref{footi}). Thus, the
stress-energy regularization is similar as for trivially scalar
fields, where the trace of $ <T_{\ab} > $ is the only piece to be
regularized necessarily. This greatly simplifies the computation.
Moreover, another very interesting (and important) property of the GRNSS
spaces is that the solution of $ T = T_1 + T_2 $ is given by $ H =
H_1 + H_2 $, where $ H_1 $, and $ H_2 $ are particular solutions
for $ T_1 $, and $ T_2 $, respectively. This last property can be
interpreted as a manifestation of the equivalence principle, since
this is the same as asserting that $ m = m_1 + m_2 $, where $ m_i
$ are the mass function in each case. Therefore, in order to find
the (well-established) quantum corrections, it is only necessary
to integrate $ G[H_{\rm q.}] = \hbox{quantum corrections} $ and
write $ H_{reg} = H_{class} + H_{q.} $ in order to know $
<T_{\ab}>$.

In fact, if one just considers (as is commonly assumed, see e.g.
\cite{bd,ip,gmm}), that semiclassical corrections are, roughly
speaking, of the form $ \propto {\cal R}^2 $, with $ {\cal R}^2 $
some geometrical Riemannian invariant of the {\em classical}
spacetime, then the corrections are obtained through direct
integration of $$  {1 \over r^2} (H_{q} r)_{,r} = \lambda {\cal
R}^2, $$ where $ \lambda $ is of order  $ G \hbar $, i.e. of order
unity in Planckian units, and is related with the number and types
of quantized fields, \cite{ip,bp}. An application of this method
to two well known cases, namely de Sitter and Schwarzschild, will
speak for itself. In the de Sitter case we have that $ R = 4 \l =
{\rm const} $, therefore $ R^2 $ ---as well as any other
Riemannian invariant--- is a constant and the quantum correction
of a macroscopic de Sitter metric turns out to be a de Sitter
term. Thus one arrives at $ H_{\rm ren} = \l_{\rm ren} r^2/3 $
(where the subscript ``ren'' means renormalized). This is a well
known result for the renormalization of a de Sitter
spacetime (see e.g. \cite{bd,gmm}). The Schwarzschild case has $ R
= 0 $, for it is a classical vacuum solution. However, $
R_{\ab\lambda\mu} R^{\ab\lambda \mu} $ is $ 48 m^2 / r^6 $. Thus,
we set a quantum vacuum contribution of the type $ \lambda m^2/r^6
$. Integration yields the remarkable result that the quantum
vacuum corrections due to its polarization yield a spacetime
metric belonging to the GRNSS spaces with $ H = 2m/r- \lambda
m^2/r^4 $. Other sources are to be found within the effective
actions of M-theory or string theory (see e.g. \cite{ayon} and references
therein\footnote{Indeed, it is not difficult to see that all those
solutions belong to the GNRSS class.}) and could be easily
incorporated into our scheme, once the corresponding Einsteinian
metric is computed. We will consider them
elsewhere.

\section{Final remarks}
\label{s-conrqibh} The first thing to be noticed is the intrinsic
freedom present in our model, which is as large as the measure of
the set of analytic functions of one variable. This is a very
rewarding feature, since it allows to impose further restrictions
coming from new proposals. In particular, it surely will be a
helpful tool when trying to find explicitly a quantum field
responsible for the $ G_{11} $ and $ G_{22} $ in the funadamental
uncharged case. For comparison, in
all previous works, based on a unique model, the prospective of
finding a quantum field related with their energy-matter
content was hopeless.

The second thing to be mentioned is, that we have here been able
to develop in depth some relevant issues left incomplete in other
works. For instance, in \cite{fmm}, the authors studied the
interpretation of the de Sitter region as a new universe birth.
They also considered the effects of the evaporation of the black
hole which would be interesting to be computed for the models
presented here. In \cite{bp}, the authors studied the stability of
the model. In the present paper, all these statements have been
substantially implemented, and this in a very natural way (see the
preceding sections).

On the other hand, the source origin for both examples in the
section before is clear, but they do not provide the right
solution of the problem we have in hand. The de Sitter one is not
enough, since a de Sitter interior cannot be directly joined with
an exterior coming from a black hole. And the Schwarzschild
solution is singular at the origin. Then, one has to find a
deviation of the de Sitter spacetime (a non-isotropic metric)
still physically meaningful, although this metric will be only
needed macroscopically. This will be the subject of subsequent
work (Refs. \cite{zn,magli2,acvv} contain hints towards this
direction).

We have here given just the first steps towards the goal of the
(semi-classical) quantization of any suitable source. Our
preliminary results show undoubtedly that this process, for the
case of \ks type of metrics, is under the same footing as when one
has to deal with a conformal field theory.

We have also shown that the solutions behave as constituting
an anomaly effect of their sources. Moreover, we have seen how
quantum field corrections can be easily incorporated into any
initial classical model within the family. A result which tells us
that the quantization scheme may be carried out more easily here
than in conformal field theory. All these findings not only compel
us to believe that black hole singularities are likely to be
removed by quantum effects but, in our view, open also a new
window for the search of a compatible quantum field that, once
regularized, may yield the same result for, at least, a particular
$ T $ inside these models.\footnote{The rotating case, which is of major
astrophysical interest, and the ---highly-- rotating and charged one,
which may be associated with spinning particles, seem to yield
results very similar to the ones presented here, see e.g. \cite{burinskii}.
This is also the outcome of preliminary calculations of ours, to be reported 
elsewhere \cite{behm}.}

Finally, the preceding sections were centered on the free field
equations. The possibility of considering some potential is
certainly important, if it can be related to some physical theory,
as in \cite{burinskii} for the case of rotating black holes. We
reckon that the field approach above may yield a different and
useful strategy towards the same goal.
\section*{Acknowledgements}
The authors acknowledge valuable discussions with A. Burinskii
(and also his careful reading of the manuscript) and with G.
Magli. This work has been supported by CICYT (Spain), project
BFM2000-0810 and by CIRIT (Generalitat de Catalunya), contracts
1999SGR-00257 and 2000ACES 00017.

\end{document}